\documentclass{amsart}
\usepackage{mathrsfs,amssymb,amsmath,amsfonts,amsxtra}
\usepackage{graphicx,color}

\begin{document}

\author{I. Schmelzer}
\thanks{Berlin, Germany}
\email{\href{mailto:ilja.schmelzer@gmail.com}{ilja.schmelzer@gmail.com}}%
\urladdr{\href{http://ilja-schmelzer.de}{ilja-schmelzer.de}}

\title[A symmetry problem in Copenhagen]{A symmetry problem in the Copenhagen interpretation} \sloppypar

\maketitle
\begin{abstract}
A non-uniqueness result for the canonical structure in quantum theory shows that the classical part of the Copenhagen interpretation contains physically important information not contained in it's quantum part. As a consequence, we cannot compute the symmetry group of a quantum theory considering only the quantum part. The unavoidable vagueness of the classical part therefore leads to a similar vagueness in the definition of the symmetry group. This makes it at least problematic, if not impossible, to establish the true symmetry group of a quantum theory in the Copenhagen interpretation. Different from the old measurement problem, the symmetry group is to important physically to be rejected as a metaphysical pseudoproblem.
\end{abstract}

\sloppypar \sloppy
\newcommand{\B}{\mbox{$\mathbb{Z}_2$}} 
\newcommand{\Z}{\mbox{$\mathbb{Z}$}}
\newcommand{\N}{\mbox{$\mathbb{N}$}}
\newcommand{\R}{\mbox{$\mathbb{R}$}}
\newcommand{\C}{\mbox{$\mathbb{C}$}}
\newcommand{\pd}{\mbox{$\partial$}}
\renewcommand{\a}{\mbox{$\hat{a}$}} 
\newcommand{\p}{\mbox{$\hat{p}$}} 
\newcommand{\q}{\mbox{$\hat{q}$}} 
\newcommand{\ps}{\mbox{$\hat{p}(s)$}} 
\newcommand{\qs}{\mbox{$\hat{q}(s)$}} 
\newcommand{\h}{\mbox{$\hat{h}$}} 

\renewcommand{\H}{\mbox{$\mathcal{H}$}} 
\renewcommand{\L}{\mbox{$\mathcal{L}^2$}} 
\renewcommand{\O}{\mbox{$\hat{\mathcal{O}}$}} 
\newcommand{\V}{\mbox{$\mathcal{V}$}} 
\providecommand{\abs}[1]{\lvert#1\rvert}

\newcommand{\Sch}{Schr\"{o}dinger\/}

\newtheorem{theorem}{Theorem}
\newtheorem{thesis}{Thesis}

\section{Introduction}

In \cite{kdv1} we have proven a non-uniqueness theorems for the canonical structure: For some fixed Hamilton operator \h, we have constructed for some real parameter $s$ different pairs \qs, \ps\/ of canonical operators so that $\h= \frac{1}{2m}\ps^2+V(\qs,s)$ with physically different, but equally nice potentials. Thus, a physically complete definition of quantum theory cannot derive the correct canonical operators \p, \q, but has to define them. Moreover, this cannot be done by pure labeling. At least the interpretation has to describe what physically distinguishes the correct canonical operators.

Discussing these consequences in \cite{kdv2}, we have found that (different from pure interpretations like the Everett \cite{Everett}, Ithaca \cite{MerminIthaca} and consistent histories \cite{Griffiths} interpretations considered there) the Copenhagen interpretation does not have a non-uniqueness problem: It is solved by the association of the correct canonical operators \p\/ and \q\/ with classically described experimental arrangements for momentum and configuration measurements. Thus, Copenhagen seems to be among the winners of this non-uniqueness result.

The aim of this paper is to show that, instead, Copenhagen is among the losers. First, we observe that an old symmetry argument made by Pauli against pilot wave theory fails now. Digging deaper, we observe that proving symmetry properties of a quantum theory in the Copenhagen interpretation appears conceptually difficult if not impossible: Once the classical part of Copenhagen solves the non-uniqueness problem, it becomes clear that this part contains physically important information. Thus, this part cannot be ignored if we consider symmetry properties of the quantum theory.

But this part is necessarily vague. How can one prove exact symmetry properties if parts of the definition of the theory are vague?  Considering some fallacious ways to solve this new symmetry problem, we observe that this problem, as well as the non-uniqueness problem, share a common structure with the good old measurement problem: One can shift the quantum-classical boundary, but all one reaches in this way is that these problems shift together with the boundary. This suggests, on the one hand, that the new symmetry problem will appear as unsolvable in the Copenhagen context as the measurement problem. On the other hand, the problem of identifying the symmetry group cannot be rejected as a metaphysical pseudoproblem as one can at least try in the case of the measurement problem.

We compare this with the situation in pilot wave and physical collapse theories. Above variants solve the non-uniqueness problem without the classical Copenhagen part, based on some additional structure, namely the configuration space $Q$ which is physically distinguished -- by some explicit trajectory or by some modification of the \Sch equation. In above cases, this violates relativistic symmetry by explicit introduction of a preferred frame.

\section{The inherent vagueness of an operationalist interpretation}\label{vagueness}

The vagueness of the classical part of the Copenhagen interpretation has been seen as a weakness. Bell's rejection of the Copenhagen interpretation as ``unprofessionally vague and ambiguous'' (\cite{Bell}, p.173) is quite famous, and many attempts to develop new interpretations of quantum theory have been motivated by the desire to get rid of this vague part, for example, Mermin ``\ldots would like to have a quantum mechanics that does not require the existence of a “classical domain”'' \cite{MerminIthaca}.

But the vagueness has been also defended as necessary, in particular by Rosenfeld. This is in particular interesting here because it illustrates the large amount of vagueness which has been defended by the Copenhagen camp:
\begin{quote}
[Everett’s] work suffers from the fundamental misunderstanding which affects all attempts at “axiomatizing” any part of physics. The “axiomatizers” do not realize that every physical theory must necessarily make use of concepts which cannot, in principle, be further analysed, since they describe the relationship between the physical system which is the object of study and the means of observation by which we study it: these concepts are those by which we give information about the experimental arrangement, enabling anyone (in principle) to repeat the experiment. It is clear that in the last resort we must here appeal to common experience as a basis for common understanding. To try (as Everett does) to include the experimental arrangement into the theoretical formalism is perfectly hopeless, since this can only shift, but never remove, this essential use of unanalysed concepts which alone makes the theory intelligible and communicable. (Leon Rosenfeld to Saul M. Bergmann, 21 Dec 1959, Rosenfeld Papers, Niels Bohr Archive, Copenhagen, as quoted by \cite{Freire})
\end{quote}
As a defense of the use of vague ``concepts which cannot, in principle, be further analysed'' it clearly fails: Nor the means of observation themself, nor their relationship with the object of the study have to be axiomatized. A fundamental theory tries to axiomatize only the fundamental constituents themself. The measurement instruments may be extremely complex and remain unanalysed, even if they consist of completely axiomatized and analysed constituents. Thus, the Rosenfeld argument, while unable to defend the vagueness, clarifies that the vagueness is unavoidable in principle in the operationalist approach, where the experimental arrangements are part of the definition of the theory.

By most working physicists, the vagueness of the classical part has not been considered as very problematic. In part because, for all practical purposes, the physical meaning of the canonical operators was really unproblematic. Then, following the invention of decoherence, it has been widely assumed that the classical part is not necessary at all, that one can derive it using decoherence techniques. If that would be true, the argument of this paper would vanish into thin air: To compute the symmetry group of the theory, it would be sufficient to compute the symmetry group of it's certain quantum part, which is essentially what has been done up to now.

But our non-uniqueness result proves that there is no such derivation, that one needs additional physical information to identify the correct choices of the canonical operators. In the Copenhagen interpretation, this additional information is given in the classical part -- by the association of the canonical operators \p, \q\/ with the experimental arrangement for their measurement. So, once different choices of \ps, \qs\/ lead to different physics, the classical part of the Copenhagen interpretation contains physically important information which cannot be derived from it's quantum part.

In the light of this observation, the vagueness of the classical part becomes dangerous. Can one be sure that the important physical information hidden behind this vagueness is unimportant for the particular physical question considered in a particular application?

\section{The Pauli argument}

One (not only) historically important case where the vagueness of the classical part has been fatal is Pauli's arguments against the pilot wave interpretation:
\begin{quote}
\ldots to ascribe $\Psi(x)$ ‘physical reality’ and not to $\phi(p)$ destroys a transformation group of the theory. (Pauli to Bohm, 03 Dec 1951, \cite{Pauli2}, 436-441, as quoted by \cite{Freire}),

\ldots the artificial asymmetry introduced in the treatment of the two variables of a canonically conjugated pair
characterizes this form of theory as artificial metaphysics. (\cite{Pauli}, as quoted by \cite{Freire}).
\end{quote}
What's wrong with it? Without doubt, there is an asymmetry between momentum and configuration already in the kinematics of pilot wave theory. But there is no symmetry between them in the kinematics of the Copenhagen intepretation too: Explicit experimental arrangements for measuring  \p\/ and \q, whatever they are, will be quite different from each other, and it is hard to imagine that they will show any symmetry. Thus, Pauli's argument is wrong, and it does not seem to be far fetched to blame the inherent, unavoidible vagueness of the classical part of Copenhagen for this error.

\section{How to establish exact symmetries of vague theories?}

This raises an important general problem: Behind the vagueness of the classical part of Copenhagen, a lot of physically important information may be hidden, and, as the Pauli precedent shows, in particular information about the symmetries of the theory. So the following question appears: \emph{How can we become sure that a Copenhagen quantum theory has some certain, exact symmetry?} Obviously one also has to check if the experimental arrangements as well as their associations with operators have this symmetry too. But these associations are only vaguely defined. This is fatal for proving exact symmetry properties: If there is no exact description of the experimental arrangement, one cannot prove that the whole theory is really symmetric.

We would not claim that it is completely impossible to argue in favour of some symmetry if the theory is vague:  In particular, one can make a good case for translational symmetries even in vague theories, simply because the meaning of our words implicitly incorporates the assumption of some translational symmetry. Nonetheless, even in this case this would be no more than a plausibility argument and not a proof: A vague description could describe a crystal as having continuous translational symmetry, while it's atomic structure has only a discrete translational symmetry. A proof would have to rely on an exact description.

If one thinks about possibilities to solve this problems, there seem to be two strategies: On the one hand, one can think that one should describe the classical part using classical physics, which would allow to prove that the measurement device follows exactly classical symmetries. On the other hand, one can think about the description of the particular experimental arrangement or at least some important part of it as being described by the quantum part, so that one can use the symmetries of the pure quantum part to prove that the experimental arrangement has the particular symmetry. Above strategies fail.

The failure of the former strategy is quite obvious -- given that we know that quantum theory is true, classical theory cannot be exact, but only an approximation. But one cannot make conclusions about exact symmetry properties based on the classical approximation. The failure of the latter strategy is more interesting.

\section{Why shifting the boundary does not solve the problem}

Desribing the measurement device or some part of it with quantum mechanics only shifts the boundary between classical and quantum part of the Copenhagen interpretation. Now, restricting the consideration to the quantum part fails, for the same reason pure quantum theory fails -- the classical part contains important physical information, and this information cannot be obtained from the pure quantum theory of some part of the world, however large. Instead, this pure quantum theory needs a further, shifted classical-quantum boundary to become physically well-defined.

It is the advantage of our non-uniqueness result that it allows to make this point certain and unquestionable: We can solve the non-uniqueness problem with Copenhagen only if we rely on some connection with some experimental arrangement, and no shift of the boundary can solve this problem: We always have to rely on some association with some external experimental arrangement, however far it has been shifted. And this association always has physical importance, because the different choices of \p\/ and \q\/ define different physics. And, as a consequence, the consideration of the symmetries of the theory is also not complete without the consideration of the next classical-quantum boundary.

\section{Three problems showing the same pattern}

This is the same problem pattern we already know from the measurement problem:  We can shift the boundary, but to solve the measurement problem we always have to rely on the (however far shifted) boundary. Thus, we have now already three problems which share this same pattern -- that we can shift the boundary as we like, but to solve these problems we have to take into account the boundary.

Up to now, only one of them, the measurement problem, has been recognized. While it has been considered as important by many of those who have studied the foundations of quantum theory, there have been also lot's of attempts to reject it as a metaphysical pseudoproblem. This is not the place to evaluate these arguments: Instead, our point is that we can now forget about them, because the new symmetry problem doesn't look like metaphysical a pseudo-problem.

The non-uniqueness problem is certainly not metaphysical:  The physical differences between the $\h= \frac{1}{2m}\ps^2+V(\qs,s)$ are far to obvious to ignore them. Of course, it does not have explicit consequences for physics as we do them: The canonical way to define quantum theory has always fixed some operators \p\/ and \q\/ and assigned them a physical meaning, thus, the ``shut up and calculate'' interpretation works as usual. It's importance is of different character: It proves that physically important information is contained in the classical-quantum boundary, information which is not contained in the quantum part taken alone, however large. And this is the key observation which makes the computation of the symmetry group problematic.

The symmetry of a physical theory is of central importance in modern physics. One can possibly argue that it's importance is   overestimated, but it is hard to deny completely it's physical importance. Thus, we obtain a new, strong physical argument against the Copenhagen interpretation:
\begin{quote}
\emph{The vagueness of the classical part of the Copenhagen interpretation prevents the exact computation of the symmetry group of the theory.}
\end{quote}
An interpretation which prevents, because of it's inherent vagueness, the computation of the exact symmetry group of a theory seems to be in deep trouble. 

It would be in deep trouble even if there would be no alternatives. But there are alternatives: Pilot wave theories \cite{Bohm} allow to derive, in quantum equilibrium, the physical predictions of quantum theory. They also allow to solve the non-uniqueness problem. On the other hand, they do not have any vague classical parts like the Copenhagen interpretation.  Everything is exactly defined, and there is no problem with computing the symmetry group of these theories. Similarly, physical collapse theories \cite{GRW,Ghirardi} are also well-defined, without reference to measurement devices, and also have no symmetry problem. Given such alternatives, what could justify a preference of an interpretation whose vagueness forbids even the computation of the symmetry group?

\section{Discussion: Does quantum theory require a preferred frame?}

The interpretations we consider as the winners of our non-uniqueness problem -- pilot wave and physical collapse theories -- share an important property: As realistic interpretations, they need a preferred frame. Does this strengthen the case for a preferred frame?  It seems. The pure interpretations we have to rejected as not viable in \cite{kdv2} have to be completed by some yet unknown additional structure, which, on the one hand, allows to prefer one \q\/ among the \qs\/ but, on the other hand, do not need for this purpose a preferred frame. This is not trivial, given that a configuration is a frame-dependent notion. In particular, in the many worlds context Wallace has noted that  ``\ldots there seems to be no relativistically covariant way to define a world \ldots'' \cite{Wallace}.

But while the currently pure interpretations may be saved by adding some appropriate structure, there seems no such way to save the Copenhagen interpretation: Once it is recognized (because of the non-uniqueness problem) that the classical part contains nontrivial physical information, and as long as this classical part remains vague, there seems no way to find out the true symmetry group of a quantum theory in this interpretation.

Thus, it seems, those who consider Lorentz symmetry as fundamental are now faced with a foundational problem: To find an interpretation of quantum theory which does not have a non-uniqueness problem, is not vague so that it does allow to compute the exact symmetry group of the theory, but does not require a preferred frame.

In our opinion, there is no need to search for such interpretations. It would have to reject the already extremely weak notion of realism which one needs to prove Bell's inequalities. Moreover, the preferred frame is not only something negative -- it also opens new ways: In particular, in \cite{clm} we have proposed a simple condensed matter interpretation for the standard model of particle physics as well as for relativistic gravity. The standard model interpretation contains lot's of important parts incompatible with a four-dimensional approach, in particular because the number of colors, the number of generations, and the number of generators of the weak group (among others) are associated with the dimension of space. These interesting possibilities are lost if we require fundamental Lorentz symmetry.

Thus, we see no good reason to bother too much about the loss of Lorentz symmetry on the fundamental level. Therefore, the existing realistic interpretations are a satisfactory way to solve the non-uniqueness problem. We prefer the de Broglie-Bohm pilot wave interpretation, because of it's mathematical beauty and because of the identity of it's predictions with standard ``shut up and calculate'' quantum theory.

Others may think that a loss of Lorentz symmetry is not permissible. They have to find now an interpretation which, on the one hand, solves the non-uniqueness problem, on the other hand, is not too vague to prevent the computation of the symmetry group. The existing pure interpretations are a good starting point, but need additional structure. Instead, the Copenhagen interpretation seems doomed -- one cannot derive exact symmetries from vague descriptions.

\end{document}